\documentclass[12pt]{article}
\usepackage{amsmath}
\usepackage{graphicx,psfrag,epsf}
\usepackage{enumerate}
\usepackage{url} 
\usepackage{float}
\usepackage{array}
\usepackage{amsmath}
\usepackage{hyperref}
\usepackage{longtable}
\usepackage{xcolor}
\usepackage{amsfonts}

\usepackage{soul}

\RequirePackage{natbib}

\newcommand{\blind}{0}

\begin{document}

\bibliographystyle{chicago}  

\def\spacingset#1{\renewcommand{\baselinestretch}%
{#1}\small\normalsize} \spacingset{1}


\if0\blind
{
  \title{\bf Superheat: An R package for creating \\beautiful and extendable heatmaps\\for visualizing complex data}
  \author{Rebecca L. Barter\\
    Department of Statistics, University of California, Berkeley\\
    and \\
    Bin Yu \\
    Department of Statistics, University of California, Berkeley}
  \maketitle
} \fi

\if1\blind
{
  \bigskip
  \bigskip
  \bigskip
  \begin{center}
    {\LARGE\bf Supervised heatmaps for visualizing complex data}
\end{center}
  \medskip
} \fi

\bigskip
\begin{abstract}
%

The technological advancements of the modern era have enabled the collection of huge amounts of data in science and beyond. Extracting useful information from such massive datasets is an ongoing challenge as traditional data visualization tools typically do not scale well in high-dimensional settings. An existing visualization technique that is particularly well suited to visualizing large datasets is the heatmap. Although heatmaps are extremely popular in fields such as bioinformatics for visualizing large gene expression datasets, they remain a severely underutilized visualization tool in modern data analysis. In this paper we introduce superheat, a new R package that provides an extremely flexible and customizable platform for visualizing large datasets using extendable heatmaps. Superheat enhances the traditional heatmap by providing a platform to visualize a wide range of data types simultaneously, adding to the heatmap a response variable as a scatterplot, model results as boxplots, correlation information as barplots, text information, and more. Superheat allows the user to explore their data to greater depths and to take advantage of the heterogeneity present in the data to inform analysis decisions. The goal of this paper is two-fold: (1) to demonstrate the potential of the heatmap as a default visualization method for a wide range of data types using reproducible examples, and (2) to highlight the customizability and ease of implementation of the superheat package in R for creating beautiful and extendable heatmaps. The capabilities and fundamental applicability of the superheat package will be explored via three case studies, each based on publicly available data sources and accompanied by a file outlining the step-by-step analytic pipeline (with code). The case studies involve (1) combining multiple sources of data to explore global organ transplantation trends and its relationship to the Human Development Index, (2) highlighting word clusters in written language using Word2Vec data, and (3) evaluating heterogeneity in the performance of a model for predicting the brain's response (as measured by fMRI) to visual stimuli.

\end{abstract}

\noindent%
\vfill

\spacingset{1.45}
\section{Introduction}
\label{sec:intro}

The rapid technological advancements of the past few decades have enabled us to collect vast amounts of data with the goal of finding answers to increasingly complex questions both in science and beyond. Although visualization has the capacity to be a powerful tool in the information extraction process of large multivariate datasets, the majority of commonly used graphical exploratory techniques such as traditional scatterplots, boxplots and histograms are embedded in spaces of 2 dimensions and rarely extend satisfactorily into higher dimensions. Basic extensions of these traditional techniques into 3 dimensions are not uncommon, but tend to be inadequately represented when compressed to a 2-dimensional format. Even graphical techniques designed for 2-dimensional visualization of multivariate data, such as the scatterplot matrix~\citep{cleveland_visualizing_1993, andrews_plots_1972} and parallel coordinates~\citep{inselberg_plane_1985, inselberg_visual_1998, inselberg_parallel_1987} become incomprehensible in the presence of too many data points or variables. These techniques suffer from a lack of scalability. Effective approaches to the visualization of high-dimensional data must subsequently satisfy a tradeoff between simplicity and complexity. A graph that is overly complex impedes comprehension, while a graph that is too simple conceals important information.\\

\noindent \emph{\large The heatmap for matrix visualization} \vspace{3mm}

\noindent An existing visualization technique that is particularly well suited to the visualization of high-dimensional multivariate data is the heatmap. At present, heatmaps are widely used in areas such as bioinformatics (often to visualize large gene expression datasets), yet are significantly underemployed in other domains~\citep{wilkinson_history_2009}. 

A heatmap can be used to visualize a data matrix by representing each matrix entry by a color corresponding to its magnitude, enabling the user to visually process large datasets with thousands of rows and/or columns. For larger matrices it is common to use clustering as a tool to group together similar observations for the purpose of revealing structure in the data and to increase the clarity of information provided by the visualization~\citep{eisen_cluster_1998}. 

Inspired by a desire to visualize a design matrix in a manner that is supervised by some response variable, we developed an R package, \textit{superheat}, for producing ``supervised'' heatmaps that extend the traditional heatmap via the incorporation of additional information. The goal of this paper is two-fold: (1) to demonstrate the potential of the heatmap as a default visualization method for data in high dimensions, and (2) to highlight the customizability and ease of implementation of the \textit{superheat} package in R for creating beautiful and extendable heatmaps.

\section{Superheat}
\label{sec:superheat}

There currently exist a number of packages in R for generating heatmaps to visualize data. The inbuilt heatmap function in R (\texttt{heatmap}) offers very little flexibility and is difficult to use to produce publication quality images. The popular visualization R package, \textit{ggplot2}, contains functions for producing visually appealing heatmaps, however \textit{ggplot2} requires the user to convert the data matrix to a long-form data frame consisting of three columns: the row index, the column index, and the corresponding fill value. Although this data structure is intuitive for other types of plots, it can be somewhat cumbersome for producing heatmaps. Some of the more recent heatmap functions, which have a focus on aesthetics are those from the \textit{pheatmap} package and its extension, \textit{aheatmap}, which allows for sample annotation. 

Our R package, \textit{superheat}, builds upon the infrastructure provided by \textit{ggplot2} to develop an intuitive heatmap function that possesses the aesthetics of \textit{ggplot2} with the simple implementation of the inbuilt heatmap functions. However it is in its customizability that superheat far surpasses all existing heatmap software. Superheat offers a number of novel features, such as smoothing of the heatmap within clusters to facilitate extremely large matrices with thousands of dimensions; overlaying the heatmap with text or numbers to increase the clarity of the data provided or to aid annotation; and most notably, the addition of scatterplots, barplots, boxplots or line plots adjacent to the rows and columns of the heatmap, allowing the user to add an entirely new layer of information. These features of superheat allow the user to explore their data to greater depths, and to take advantage of the heterogeneity present in the data to inform analysis decisions.

Although there are a vast number of potential use cases of superheat for data exploration, in this paper we will present three case studies that highlight the features of superheat for (1) combining multiple sources of data together, (2) uncovering correlational structure in data, and (3) evaluating heterogeneity in the performance of data models.

\spacingset{1.45} 
\section{Case study I: combining data sources to explore global organ transplantation trends}
\label{sec:transplant}

The demand worldwide for organ transplantation has drastically increased over the past decade, leading to a gross imbalance of supply and demand. For example, in the United States, there are currently over 100,000 people waiting on the national transplant list but there simply aren't enough donors to meet this demand \citep{abouna_organ_2008}. Unfortunately this problem is worse in some countries than others as organ donation rates vary hugely from country to country, and it has been suggested that organ donation and transplantation rates are correlated with country development \citep{garcia_global_2012}. This case study will explore combining multiple sources of data in order to examine the recent trends in organ donation worldwide as well as the relationship between organ donation and the Human Development Index (HDI). 

The organ donation data was collected from the WHO-ONT Global Observatory on Donation and Transplantation, which represents the most comprehensive source to date of worldwide data concerning activities in organ donation and transplantation derived from official sources. The database (available at \url{http://www.transplant-observatory.org/export-database/}) contains information from a questionnaire annually distributed to health authorities from the 194 Member States in the six World Health Organization (WHO) regions: Africa, The Americas, Eastern Mediterranean, Europe, South-East Asia and Western Pacific. 

Meanwhile, the HDI was created to emphasize that people and their capabilities (rather than economic growth) should be the ultimate criteria for assessing the development of a country. The HDI is calculated based on life expectancy, education and per capita indicators and is hosted by the United Nations Development Program's Human Development Reports (available at \url{http://hdr.undp.org/en/data#}).

\subsection{Exploration}

\begin{figure}[H]
\centering
\includegraphics[scale = 0.41]{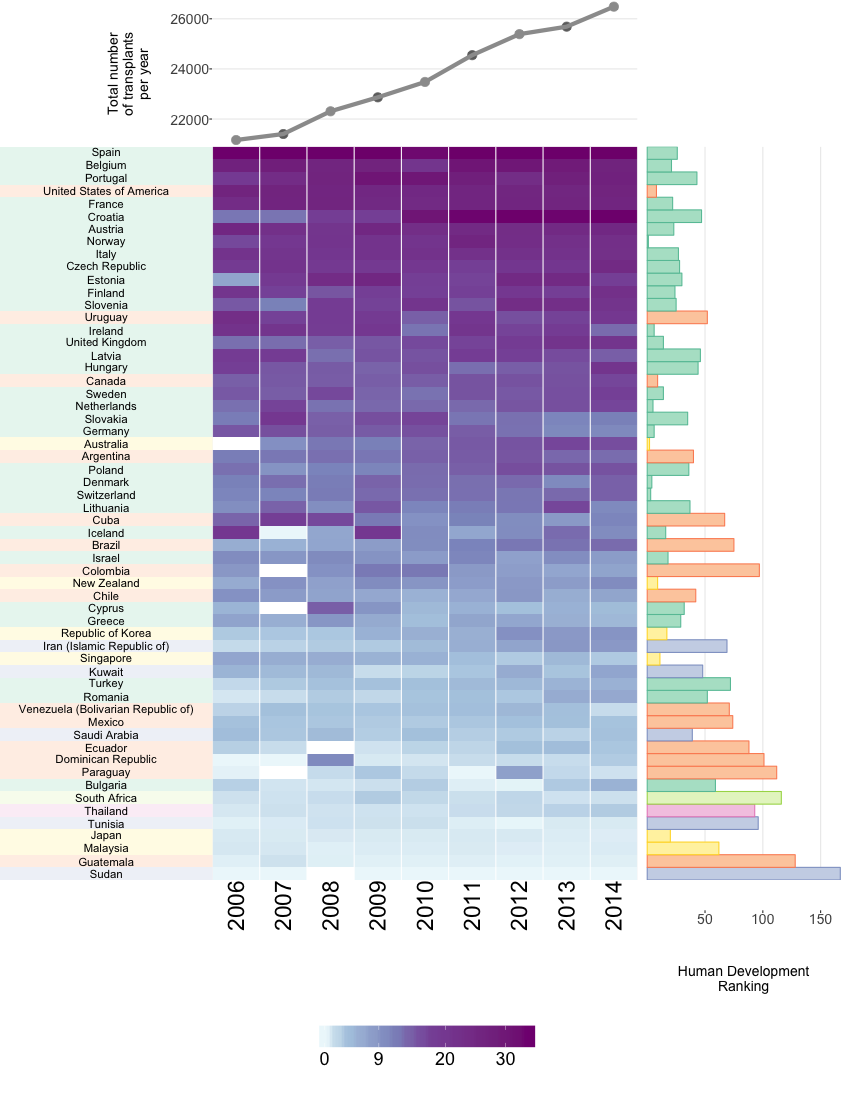}
\caption{A superheatmap examining organ donations per country and its relation to HDI. The bar plot to the right of the heatmap displays the HDI ranking (a lower ranking is better). Each cell in the heatmap is colored based on the number of organ donations from deceased donors per 100,000 individuals for the corresponding country and year. White cells correspond to missing values. The rows (countries) are ordered by the average number of transplants per 100,000 (averaging across years). The country labels and HDI bar plot are colored based on which region the country belongs to: Europe (green), Eastern Mediterranean (purple), Western Pacific (yellow), America (orange), South East Asia (pink) and Africa (light green). The line plot above the heatmap displays the aggregate total number of organs donated per year. }
\label{fig:organ}
\end{figure}

In the superheatmap presented in Figure \ref{fig:organ}, the central heatmap presents the total number of donated organs from deceased donors per 100,000 individuals between 2006 to 2014 for each country (restricting to countries for which data was collected for at least 8 of the 9 years). 

Above the heatmap, a line plot displays the overall number of donated organs over time, aggregated across all 58 countries represented in the figure. We see that overall, the organ donation rate is increasing, with approximately 5,000 more recorded organ donations occurring in 2014 relative to 2006. To the right of the heatmap, next to each row, a bar displays the country's HDI ranking. Each country is colored based on which global region it belongs to: Europe (green), Eastern Mediterranean (purple), Western Pacific (yellow), America (orange), South East Asia (pink) and Africa (light green).

From Figure \ref{fig:organ}, we see that Spain is the clear leader in global organ donation, however there has been a rapid increase in donation rates in Croatia, which had one of the lower rates of organ donation in 2006 but has a rate equaling that of Spain in 2014. However, in contrast to the growth experienced by Croatia, the rate of organ donation appears to be slowing in several countries including as Germany, Slovakia and Cuba. For some unexplained reason, Iceland had no organ donations recorded from deceased donors in 2007.

The countries with the most organ donations are predominantly European and American. In addition, there appears to be a general correlation between organ donations and HDI ranking: countries with lower (better) HDI rankings tend to have higher organ donation rates. Subsequently, countries with higher (worse) HDI rankings tend to have lower organ donation rates, with the exception of a few Western Pacific countries such as Japan, Singapore and Korea, which have fairly good HDI rankings but relatively low organ donation rates.

In this case study, superheat allowed us to visualize multiple trends simultaneously without resorting to mass overplotting. In particular, we were able to examine the organ donation over time and for each country and compare these trends to the country's HDI ranking while visually grouping countries from the same region together. No other 2-dimensional graph would be able to provide such an in-depth, yet uncluttered, summary of the trends contained in these data.

The code used to produce Figure \ref{fig:organ} is provided in Appendix \ref{sec:code}, and a summary of the entire analytic pipeline is provided in the supplementary materials as well as at: \url{https://rlbarter.github.io/superheat-examples/}.

\spacingset{1.45} 
\section{Case study II: uncovering clusters in language using Word2Vec}
\label{sec:Word2Vec}

Word2Vec is an extremely popular group of algorithms for embedding words into high-dimensional spaces such that their relative distances to one another convey semantic meaning \citep{mikolov_efficient_2013}. The canonical example highlighting the impressiveness of these word embeddings is
$$\overrightarrow{\text{man}} - \overrightarrow{\text{king}} + \overrightarrow{\text{woman}} = \overrightarrow{\text{queen}}.$$

That is, that if you take the word vector for ``man'', subtract the word vector for ``king'' and add the word vector for ``woman'', you approximately arrive at the word vector for ``queen''. These algorithms are quite remarkable and represent an exciting step towards teaching machines to understand language.

In 2013, Google published pre-trained vectors trained on part of the Google News corpus, which consists of around 100 billion words. Their algorithm produced 300-dimensional vectors for 3 million words and phrases (the data is hosted at \url{https://code.google.com/archive/p/word2vec/}).

The majority of existing visualization methods for word vectors focus on projecting the 300-dimensional space to a low-dimensional representation using methods such as t-distributed stochastic neighbor embedding (t-SNE) \citep{maaten_visualizing_2008}.

\subsection{Visualizing cosine similarity}

\begin{figure}[H]
\centering
\includegraphics[scale = 0.59]{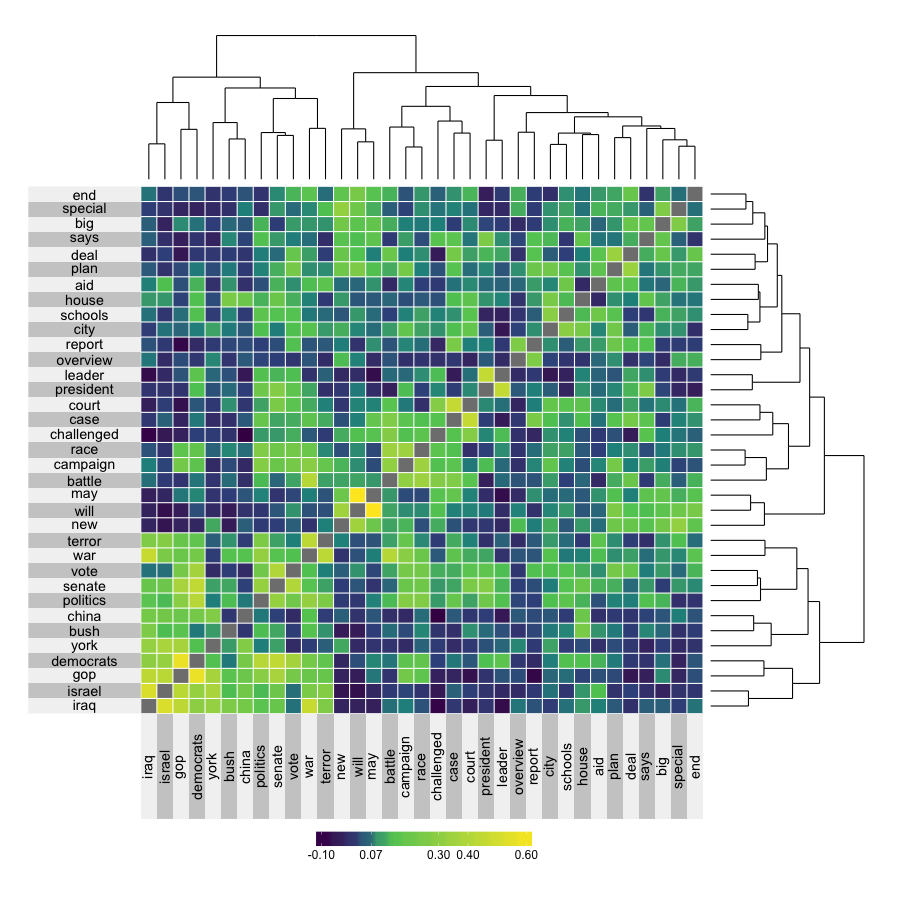}
\caption{The cosine similarity matrix for the 35 most common words from the NY Times headlines that also appear in the Google News corpus. The rows and columns are ordered based on hierarchical clustering. This hierarchical clustering is displayed via dendrograms.}
\label{fig:Word2Vec}
\end{figure}

In this superheat case study we present an alternative approach to visualizing word vectors, which highlights contextual similarity. Figure \ref{fig:Word2Vec} presents the cosine similarity matrix for the GoogleNews word vectors of the 35 most common words from the NY Times headlines dataset (from the \textit{RTextTools} package). The rows and columns are ordered based on a hierarchical clustering and are accompanied by dendrograms describing this hierarchical cluster structure. From this superheatmap we observe that words appearing in global conflict contexts, such as ``terror'' and ``war'', have high cosine similarity (implying that these words appear in similar contexts). Words that are used in legal contexts, such as ``court'' and ``case'', as well as words with political context such as ``Democrats'' and ``GOP'' also have high pairwise cosine similarity.

The code used to present Figure \ref{fig:Word2Vec} is provided in Appendix \ref{sec:code}.

\newpage

Although the example presented in Figure \ref{fig:Word2Vec} displays relatively few words (we are presenting only the 35 most frequent words) and we have reached our capacity to be able to visualize each word individually on a single page, it is possible to use superheat to represent hundreds or thousands of words simultaneously by aggregating over word clusters.

\subsection{Visualizing word clusters}

\begin{figure}[H]
\centering
\includegraphics[scale = 0.3]{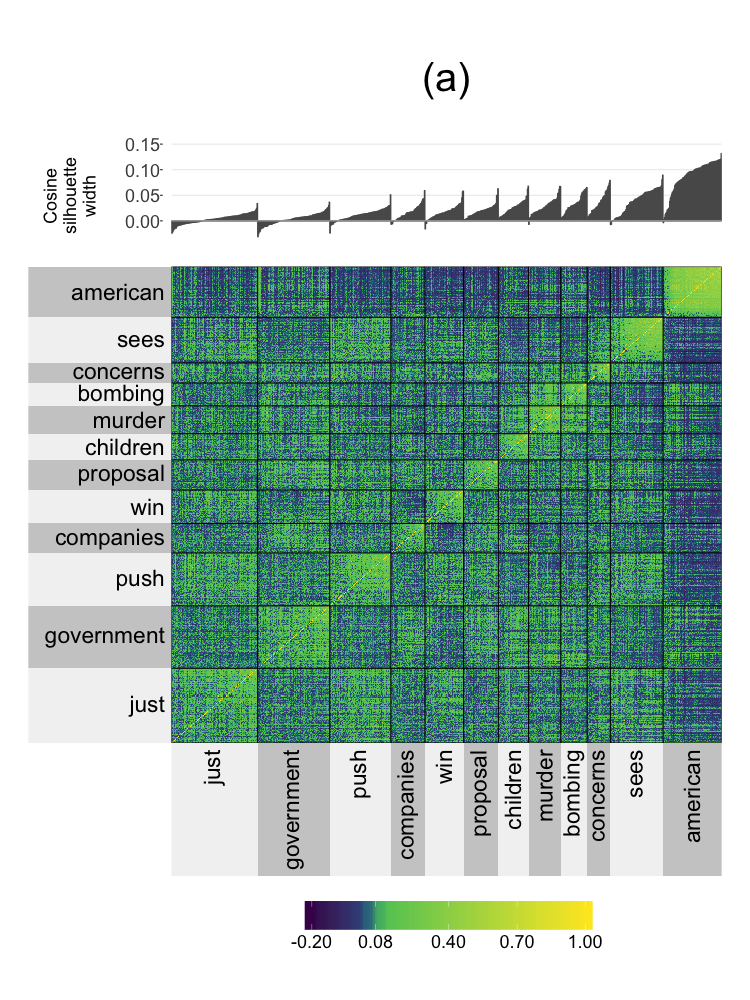}
\includegraphics[scale = 0.3]{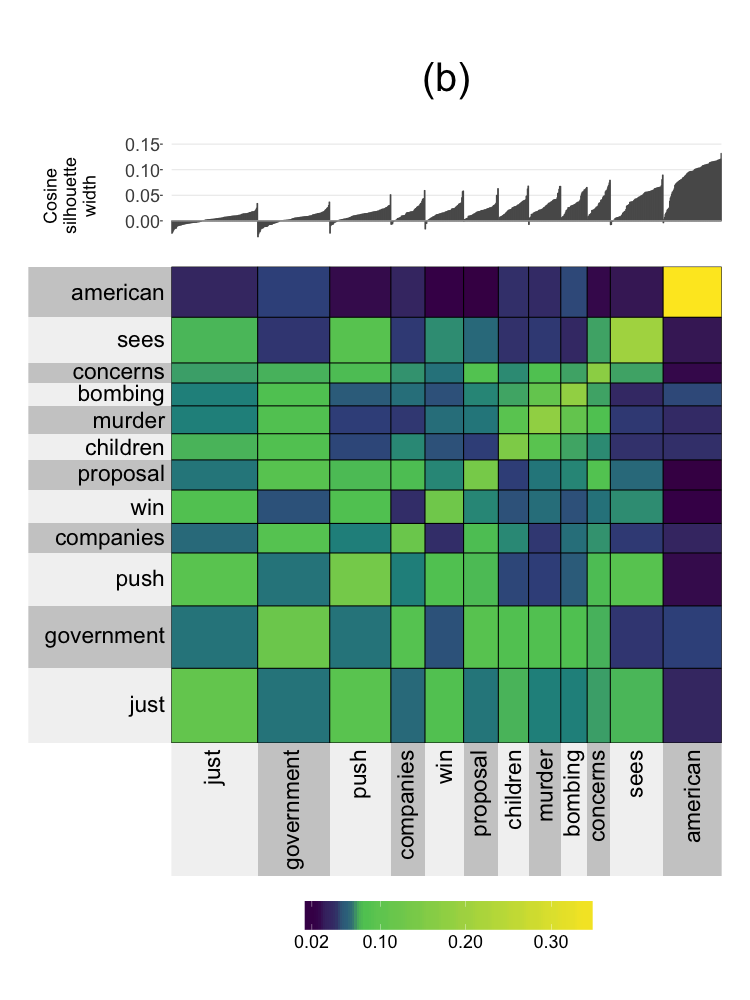}
\caption{A clustered cosine similarity matrix for the 855 most common words from the NY Times headlines that also appear in the Google News corpus. The clusters were generated using PAM and the cluster label is given by the medoid word of the cluster. Panel (a) displays the raw clustered $855 \times 855$ cosine similarity matrix, while panel (b) displays a ``smoothed'' version where the cells in the cluster are aggregated by taking the median of the values within the cluster.}
\label{fig:Word2Vec-cluster}
\end{figure}

Figure \ref{fig:Word2Vec-cluster}(a) displays the cosine similarity matrix for the Google News word vectors of the 855 most common words from the NY Times headlines dataset where the words are grouped into 12 clusters generated using the Partitioning Around Medoids (PAM) algorithm \citep{kaufman_partitioning_1990, reynolds_clustering_2006} applied to the rows/columns of the cosine similarity matrix. As PAM forces the cluster centroids to be data points, we represent each cluster by the word that corresponds to its center (these are the row and column labels that appear in Figure \ref{fig:Word2Vec-cluster}(a)).  A silhouette plot is placed above the columns of the superheatmap in Figure \ref{fig:Word2Vec-cluster}(a), and the clusters are ordered in increasing average silhouette width.

The number of clusters ($k = 12$) was chosen based on the value of $k$ that was optimal based on two types of criteria: (1) performance-based \citep{rousseeuw_silhouettes:_1987}: the maximal average cosine-silhouette width, and (2) stability-based \citep{yu_stability_2013}: the average pairwise Jaccard similarity based on 100 membership vectors each generated by a 90\% subsample of the data. Plots of $k$ versus average silhouette width and average Jaccard similarity are presented in Appendix \ref{sec:pam}. The silhouette width is a traditional measure of cluster quality based on how well each object lies within its cluster, however we adapted its definition to suit cosine-based distance so that the cosine-silhouette width for data point $i$ is defined to be:

$$sil_{\text{cosine}}(i) = b(i) - a(i)$$

\noindent where  $a(i) = \frac{1}{\|C_i\|} \sum_{j \in C_i} d_{\text{cosine}}(x_i, x_j)$ is the average cosine-dissimilarity of $i$ with all other data within the same cluster ($C_i$ is the index set of the cluster to which $i$ belongs), and $b(i) = \min_{C \neq C_i}  d_{\text{cosine}}(x_i, C)$ is the lowest average dissimilarity of $i$ to any other cluster of which $i$ is not a member. $d_{\text{cosine}}(x, y)$ is a measure of cosine ``distance'', which is equal to $d_{\text{cosine}} = \frac{cos^{-1}(s_{\text{cosine}})}{\pi}$ (where $s_{\text{cosine}}$ is standard cosine similarity).

Word clouds displaying the words that are members of each of the 12 word clusters are presented in Appendix \ref{sec:wordcloud}. For example, the ``government'' cluster contains words that typically appear in political contexts such as ``president'', ``leader'', and ``senate'', whereas the ``children'' cluster contains words such as ``schools'', ``home'', and ``family''.

Figure \ref{fig:Word2Vec-cluster}(b) presents a ``smoothed'' version of the cosine similarity matrix in panel (a), wherein the smoothed cluster-aggregated value corresponds to the median of the original values in the original ``un-smoothed'' matrix. The smoothing provides an aggregated representation of Figure \ref{fig:Word2Vec-cluster}(a) that allows the viewer to focus on the overall differences between the clusters. Note that the color range is slightly different between panels (a) and (b) due to the extreme values present in panel (a) being removed when we take the median in panel (b).

What we find is that the words in the ``American'' cluster have high silhouette widths, and thus is a ``tight'' cluster. This is reflected in the high cosine similarity within the cluster and low similarity between the words in the ``American'' cluster and words from other clusters. However, the words in the ``murder'' cluster have relatively high cosine similarity with words in the ``government'', ``children'', ``bombing`` and ``concerns'' clusters. The clusters whose centers are not topic-specific such as ``just'', ``push'', and ``sees'' tend to consist of common words that are context agnostic (see their word clouds in Appendix \ref{sec:wordcloud}), and these clusters have fairly high average similarity with one another.

The information presented by Figure \ref{fig:Word2Vec-cluster} far surpasses that of a standard silhouette plot: it allows the quality of the clusters to be evaluated relative to one another. For example, when a cluster exhibits low between-cluster separability, we can clearly see \textit{which} clusters it is close to.

The code used to produce Figure \ref{fig:Word2Vec-cluster} is provided in Appendix \ref{sec:code}, and a summary of the entire analytic pipeline is provided in the supplementary materials as well as at: \url{https://rlbarter.github.io/superheat-examples/}.

\spacingset{1.45} 
\section{Case study III: evaluation of heterogeneity in the performance of predictive models for fMRI brain signals from image inputs}
\label{sec:fmri}

Our final case study evaluates the performance of a number of models of the brain's response to visual stimuli. This study is based on data collected from a functional Magnetic Resonance Imaging (fMRI) experiment performed on a single individual by the Gallant neuroscience lab at UC Berkeley~\citep{vu_nonparametric_2009, vu_encoding_2011}. fMRI measures oxygenated blood flow in the brain, which can be considered as an indirect measure of neural activity (the two processes are highly correlated). The measurements obtained from an fMRI experiment correspond to the aggregated response of hundreds of thousands of neurons within cube-like voxels of the brain, where the segmentation of the brain into 3D voxels is analogous to the segmentation of an image into 2D pixels.

The data contains the fMRI measurements (averaged over 10 runs of the experiment) for each of 1,294 voxels located in the V1 region of the visual cortex of a single individual in response to viewings of 1,750 different images (such as a picture of a baby, a house or a horse).

\begin{figure}[H]
\begin{center}
\includegraphics[scale = 0.48]{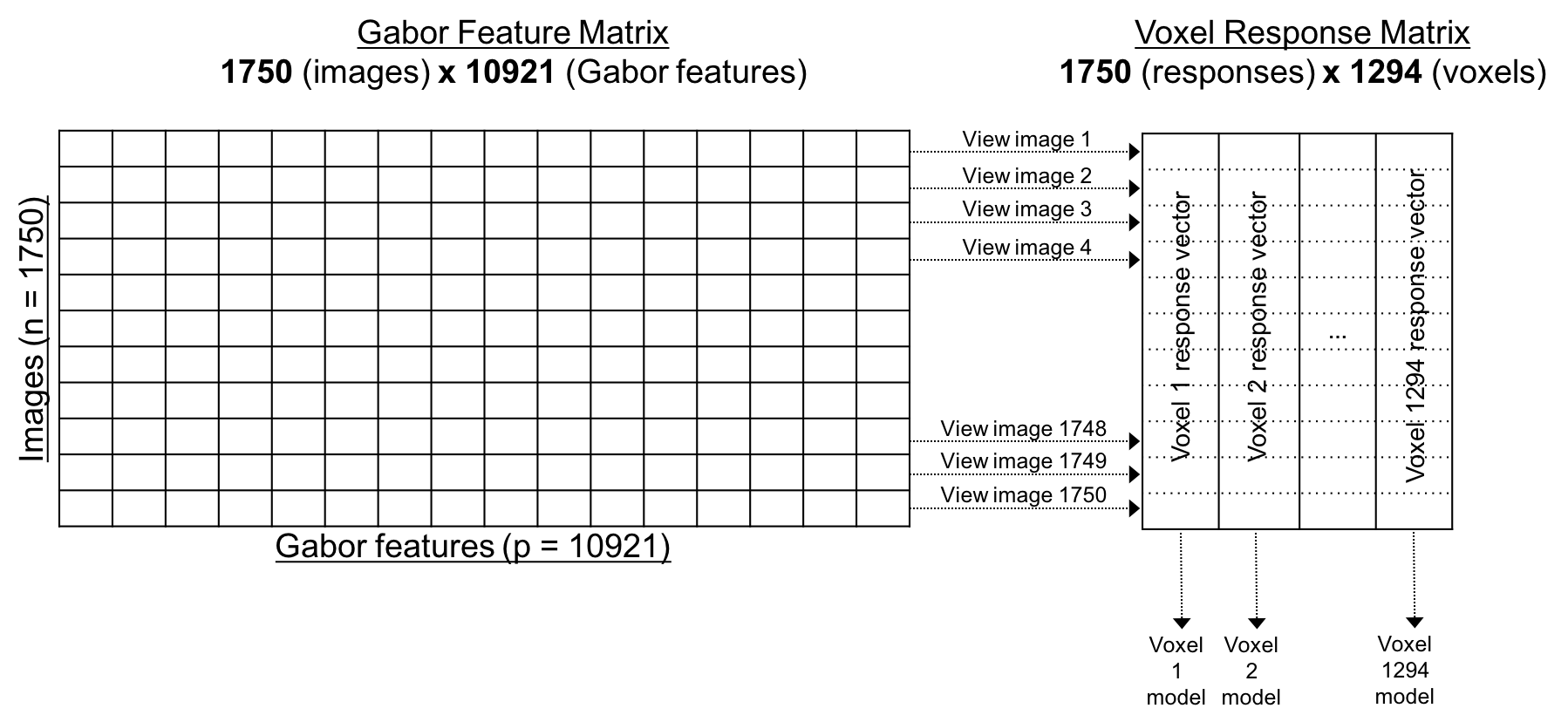}
\end{center}
\caption{A diagram describing the fMRI data containing (1) a design matrix with 1,750 observations (images) and 10,921 features (Gabor wavelets) for each image, and (2) a voxel response matrix consisting of 1,294 distinct voxel response vectors where for each voxel the responses to each of the 1,750 images were recorded. We fit a predictive model for each voxel using the Gabor feature matrix (so that we have 1,294 models). The heatmap in Figure~\ref{fig:all-voxel} corresponds to the voxel response matrix.}
\label{fig:fMRIdat}
\end{figure}

Each image is a $128 \times 128$ pixel grayscale image, which is represented as a vector of length $10,921$ through a Gabor wavelet transformation~\citep{lee_image_1996}. 

The data are hosted on the Collaborative Research in Computational Neuroscience repository (and can be found at \url{https://crcns.org/data-sets/vc/vim-1}). However, unfortunately, only the voxel responses and raw images are available. The Gabor wavelet features are not provided. Figure~\ref{fig:fMRIdat} displays a graphical representation of the data structure.

\subsection{Modeling brain activity}

We developed a model for each voxel that predicts its response to visual stimuli in the form of greyscale images. Since each voxel responds quite differently to the image stimuli, instead of fitting a single multi-response model, we fit 1,294 independent Lasso models as in \cite{vu_encoding_2011}. The models are then evaluated based on how well they predict the voxel responses to a set of 120 withheld validation images.

\begin{figure}[H]
\begin{center}
\includegraphics[scale = 0.28]{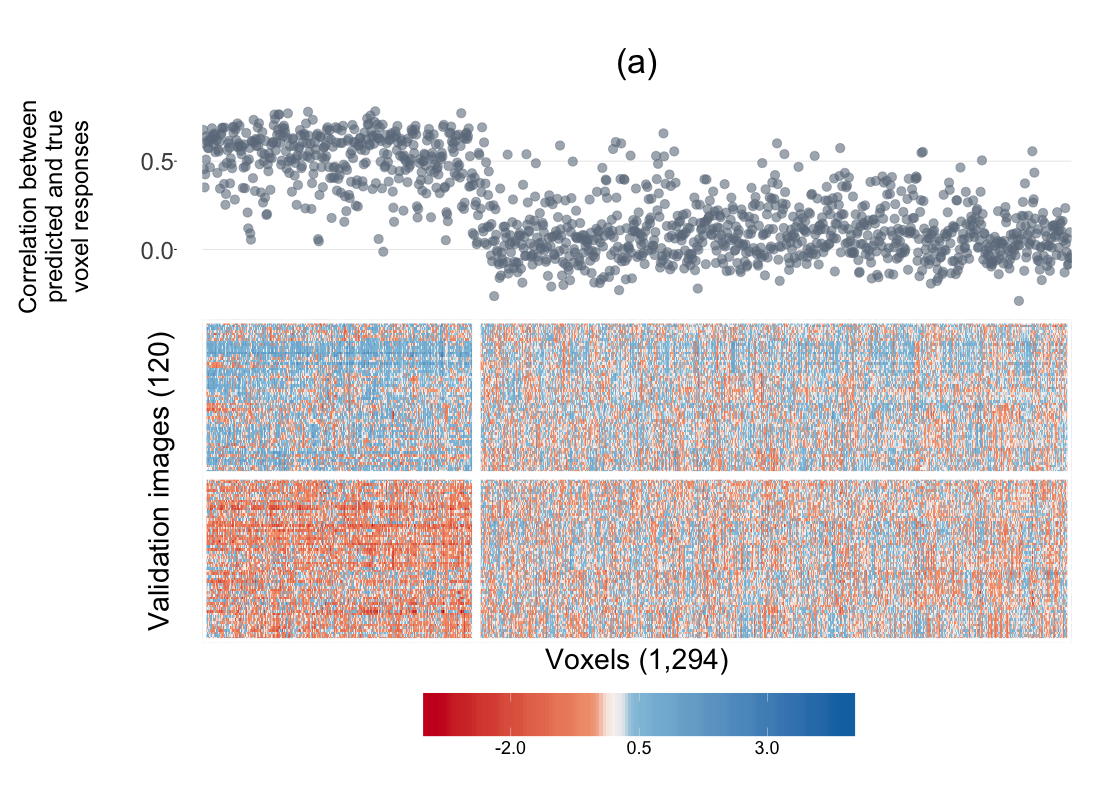}
\includegraphics[scale = 0.28]{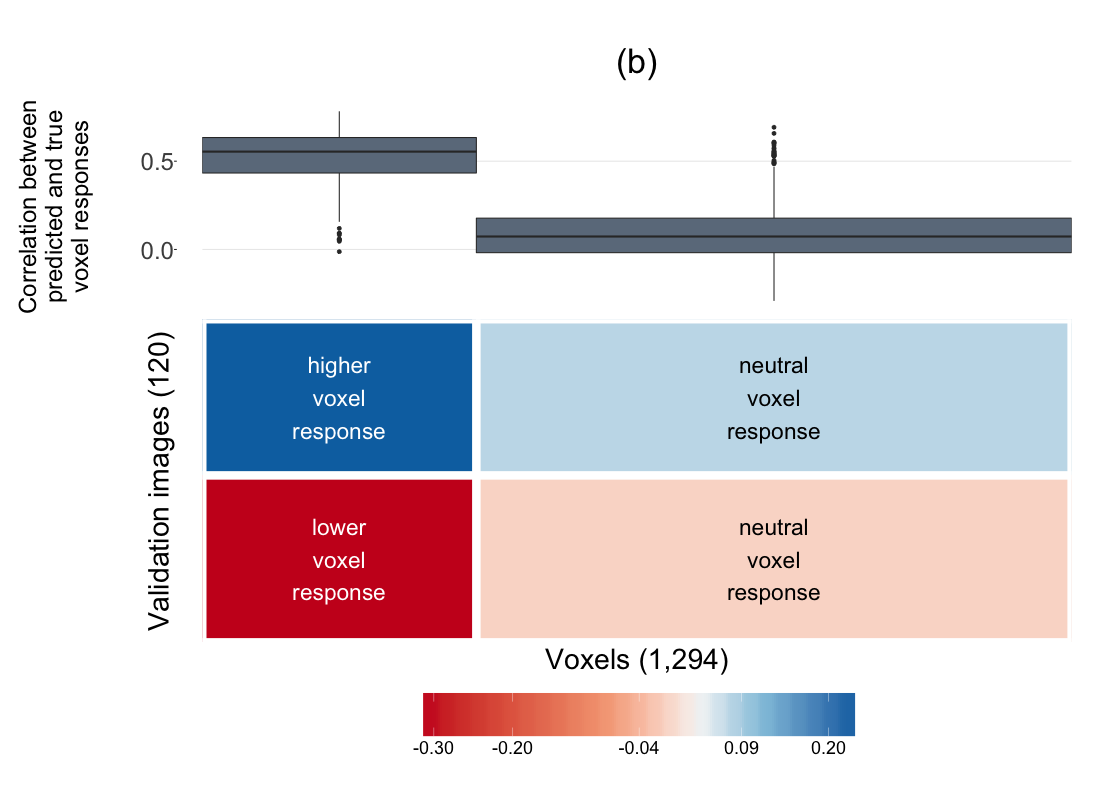}
\end{center}
\caption{A superheatmap displaying the validation set voxel response matrix (Panel (a) displays the raw matrix, while Panel (b) displays a smoothed version). The images (rows) and voxels (columns) are each clustered into two groups (using K-means). The left cluster of voxels are more ``sensitive'' wherein their response is different for each group of images (higher than the average response for top cluster images, and lower than the average response for bottom cluster images), while the right cluster of voxels are more ``neutral'' wherein their response is similar for both image clusters. Voxel-specific Lasso model performance is plotted as correlations above the columns of the heatmap (as a scatterplot in (a) and cluster-aggregated boxplots in (b)).}
\label{fig:all-voxel}
\end{figure}

\subsection{Simultaneous performance evaluation of all 1,294 voxel-models}

The voxel response matrix is displayed in Figure \ref{fig:all-voxel}(a). The rows of the heatmap correspond to the 120 images from the validation set, while the columns correspond to the 1,294 voxels. Each cell displays to the voxel's response to the image. The rows and columns are clustered into two groups using K-means. As in Figure \ref{fig:Word2Vec-cluster}(a), the heatmap is extremely grainy. Figure \ref{fig:all-voxel}(b) displays the same heatmap with the cell values smoothed within each cluster (by taking the median value). 

Appendix \ref{sec:images} displays four randomly selected images from each of the two image clusters. We find that the bottom image cluster consists of images for which the subject is easily identifiable (e.g. Princess Diana and Prince Charles riding in a carriage, a bird, or an insect), whereas the contents of images from the top cluster of images are less easy to identify (e.g. rocks, a bunch of apples, or an abstract painting). Further, from Figure \ref{fig:all-voxel}, it is clear that the brain is much more active in response to the images from the top cluster (whose contents were less easily identifiable) than to images from the bottom cluster.

Furthermore, there are two distinct groups of voxels:

\begin{enumerate}
\item \textbf{Sensitive voxels} that respond very differently to the two groups of images (for the top image cluster, their response is significantly lower than the average response, while for the bottom image cluster, their response is significantly higher than the average response).
\item \textbf{Neutral voxels} that respond similarly to both clusters of images.
\end{enumerate}
In addition, above each voxel (column) in the heatmap, the correlation of that voxel-model's predicted responses with the voxel's true response is presented (as a scatterplot in Panel (a) and as aggregate boxplots in Panel (b)). It is clear that the models for the voxels in the first (sensitive) cluster perform significantly better than the models for the voxels in the second (neutral) cluster. That is, the responses of the voxels that are sensitive to the image stimuli are much easier to predict (the average correlation between the predicted and true responses was greater than 0.5) than the responses of the voxels whose responses are neutral (the average correlation between the predicted and true responses was close to zero). 

Further examination revealed that the neutral voxels were primarily located on the periphery of the V1 region of the visual cortex, whereas the sensitive voxels tended to be more centrally located.

Although a standard histogram of the predicted and observed response correlations would have revealed that there were two groups of voxels (those whose responses we can predict well, and those whose responses we cannot), superheat allowed us to examine this finding in context. In particular, it allowed us to take advantage of the heterogeneity present in the data: we were able to identify that the voxels whose response we were able to predict well were exactly the voxels whose response was sensitive to the two clusters of images.

Note that we also ran Random Forest models for predicting the voxel responses and found the same results, however, the overall correlation was approximately 0.05 higher on average.

The code used to produce Figure \ref{fig:all-voxel} is provided in Appendix \ref{sec:code}, and a summary of the entire analytic pipeline is provided in the supplementary materials as well as at: \url{https://rlbarter.github.io/superheat-examples/}.

\spacingset{1.45} 
\section{Implementation of supervised heatmaps}
\label{sec:R}

As evident from the examples presented above, superheatmaps are flexible, customizable and very useful. Such plots would be difficult and time-consuming to produce without the existence of software that can automatically generate the plots given the user's preferences. The \textit{superheat} software package written by the authors implements superheatmaps in the \textit{R} programming language. The package makes use of the popular \textit{ggplot2} package, but does not utilize the \textit{ggplot2} grammar of graphics~\citep{wickham_layered_2010}. In particular, \textit{superheat} exists as a stand-alone function with the sole purpose of producing customizable supervised heatmaps. The development page for \textit{superheat} is hosted openly at \url{https://github.com/rlbarter/superheat}, where the user can also find a detailed Vignette (\url{https://rlbarter.github.io/superheat/}) describing further information on the specific usage of the plot in R as well as a host of options for customizability. Details of the analytic pipeline for the case studies presented in this paper can be found in the supplementary materials as well as at \url{https://rlbarter.github.io/superheat-examples/}. See Appendix~\ref{sec:code} for the example code that produced the superheatmaps appearing in this paper.

There are two main types of clustering customization for the \textit{superheat} package. The user has the option of providing their own clustering algorithm as a predefined membership vector, or the user can simply specify how many clusters they would like. The default clustering algorithm is K-means. To select the number of clusters, it is recommended that the user does so prior to the implementation of the supervised heatmaps using standard methods such as Silhouette plots~\citep{rousseeuw_silhouettes:_1987}.

A vast number of aesthetic options exist for the supervised heatmaps. For instance, each of the figures presented in this paper exhibited unique color schemes. Customizability of the color scheme is possible, not only for the color scale in the the heatmap, but also for the adjacent plots, row and column labels, grid lines, and more. The default color scheme of superheat is the perceptually uniform viridis colormap (this was the color scheme used for the Word2Vec Case Study II). 

Moreover, there are several options for adjacent plots: they can be scatterplots (the default), scatterplots with a smoothed curve, an isolated smoothed curve, barplots, line plots, scatterplots with points connected by lines, and dendrograms. These options, and more, are demonstrated in the Vignette that can be found at \url{https://rlbarter.github.io/superheat/}.

\spacingset{1.45} 
\section{Conclusion}
\label{sec:conclusion}

In this paper, we have proposed the superheatmap that aguments traditional heatmaps primarily via the inclusion of extra information such as a response variable as a scatterplot, model results as boxplots, correlation information as barplots, text information, and more. These augmentations provide the user with an additional avenue for information extraction, and allow for exploration of heterogeneity within the data. The superheatmap, as implemented by the \textit{superheat} package written by the authors, is highly customizable and can be used effectively in a wide range of situations in exploratory data analysis and model assessment. The usefulness of the supervised heatmap was highlighted in three case studies. The first combined multiple sources of data to assess the relationship between organ donation and country development worldwide. The second explored the structure of the english language by visualizing word clusters from Word2Vec data, while highlighting the hierarchical nature of these word groupings. Finally, the third case study evaluated heterogeneity in the performance of Lasso models designed to predict fMRI brain signals in response to visual stimuli in the form of image viewings. We hope that we have demonstrated clearly that
the heatmap is an extremely useful data visualization tool, particularly for high-dimensional datasets.

\newpage
\appendix
\section{APPENDIX: IMPLEMENTATION IN R}
\label{sec:code}
The \textit{superheat} package introduces a new function, conveniently also named \texttt{superheat}. The use of this function to generates the superheat Figures in this paper is shown below. The code for the complete analysis pipeline for these case studies can be found in the supplementary materials as well as at \url{https://rlbarter.github.io/superheat-examples/}.

Below we display the code for Figure \ref{fig:organ} (global organ donation and human development).
\begin{verbatim}
superheat(donor.matrix,
    # set heatmap color map
          heat.pal = brewer.pal(5, "BuPu"),
          heat.na.col = "white",
    # order rows in increasing order of donations
          order.rows = order(organs.by.country),
    # grid line colors
          grid.vline.col = "white",
    # right plot: HDI
          yr = hdi.match.2014$rank,
          yr.plot.type = "bar",
          yr.axis.name = "Human Development Ranking",
          yr.bar.col = region.col.dark,
          yr.obs.col = region.col,
    # top plot: donations by year
          yt = organs.by.year,
          yt.plot.type = "scatterline",
          yt.axis.name = "Total number of transplants per year",    
    # left labels
          left.label.col = adjustcolor(region.col, alpha.f = 0.3),
    # bottom labels
          bottom.label.col = "white",
          bottom.label.text.angle = 90,
          bottom.label.text.alignment = "right")
\end{verbatim}

\newpage

Next we display the superheat code to produce Figure \ref{fig:Word2Vec} (the cosine similarity matrix for the 35 most common words from the NY Times headlines).
\begin{verbatim}
superheat(cosine.similarity, 
     # dendrograms
          row.dendrogram = T, 
          col.dendrogram = T,     
     # gird lines
          grid.hline.col = "white",
          grid.vline.col = "white",
     # bottom label
          bottom.label.text.angle = 90)
\end{verbatim}
\newpage

The following displays the code to produce Figure \ref{fig:Word2Vec-cluster} (the clustered cosine similarity matrix for the 855 most common words from the NY Times headlines). To produce the smoothed version of the superheatmap, Panel (b), we use the argument \texttt{smooth.heat = TRUE}.
\begin{verbatim}
superheat(cosine.similarity.full,
   # cluster membership for words          
          membership.rows = word.membership,
          membership.cols = word.membership,
    # silhouette plot
          yt = cosine.silhouette.width,
          yt.axis.name = "Cosine silhouette width",
          yt.plot.type = "bar",
          yt.bar.col = "grey35",
    # row/col order
          order.rows = order(cosine.silhouette.width),
          order.cols = order(cosine.silhouette.width),
    # labels
          bottom.label.text.angle = 90,
          bottom.label.text.alignment = "right",
          left.label.text.alignment = "right",
     # smoothing option
          smooth.heat = TRUE)
\end{verbatim}
\newpage

Finally we display the code to produce Figure \ref{fig:all-voxel}: the validation set voxel response matrix and Lasso model performance. To produce the smoothed version of the superheatmap, Panel (b), we use the arguments \texttt{smooth.heat = TRUE} and \texttt{yt.plot.type = "boxplot"}.
\begin{verbatim}
superheat(val.resp, 
    # color scheme          
          heat.pal = brewer.pal(5, "RdBu"),
    # row and column clustering
          membership.rows = image.clusters,
          membership.cols = voxel.clusters,
    # top plot
          yt = prediction.cor$cor,
          yt.axis.name = "Correlation between predicted and true voxel responses",
          yt.obs.col = rep("slategray4", ncol(val.resp)),
          yt.point.alpha = 0.6,
    # labels   
          left.label = "none",
          bottom.label = "none",
    # grid lines
          grid.hline.col = "white",
          grid.vline.col = "white",
    # row and column titles
          row.title = "Validation images (120)",
          column.title = "Voxels (1,294)")
\end{verbatim}
\newpage

\section{APPENDIX: Selecting the number of word clusters}
\label{sec:pam}

\begin{figure}[H]
\centering
\includegraphics[scale = 0.15]{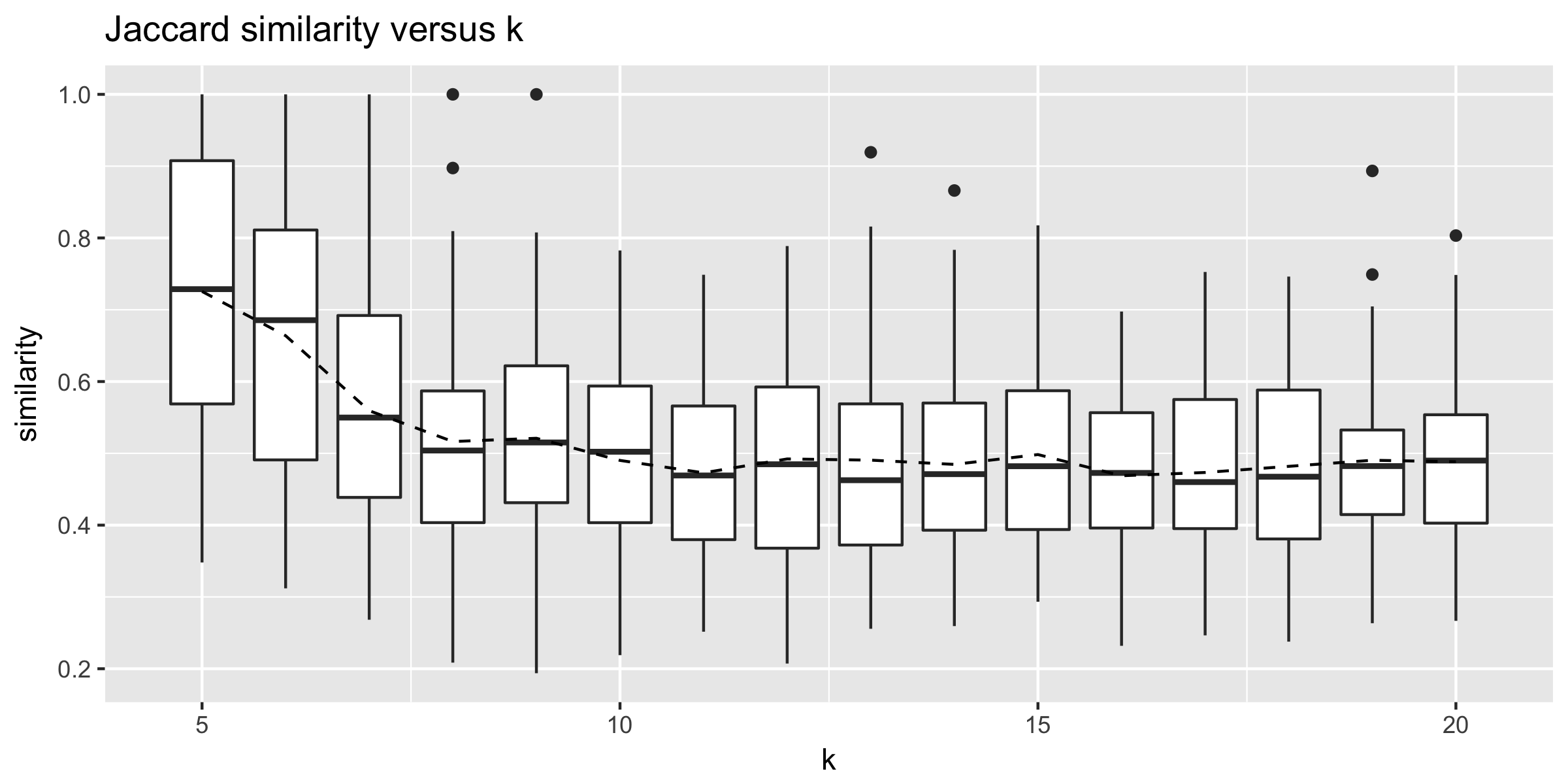}
\caption{Average pairwise Jaccard Similarity between 100 90\% subsamples of the set of word vectors.}
\label{fig:jaccard}
\end{figure}

\begin{figure}[H]
\centering
\includegraphics[scale = 0.15]{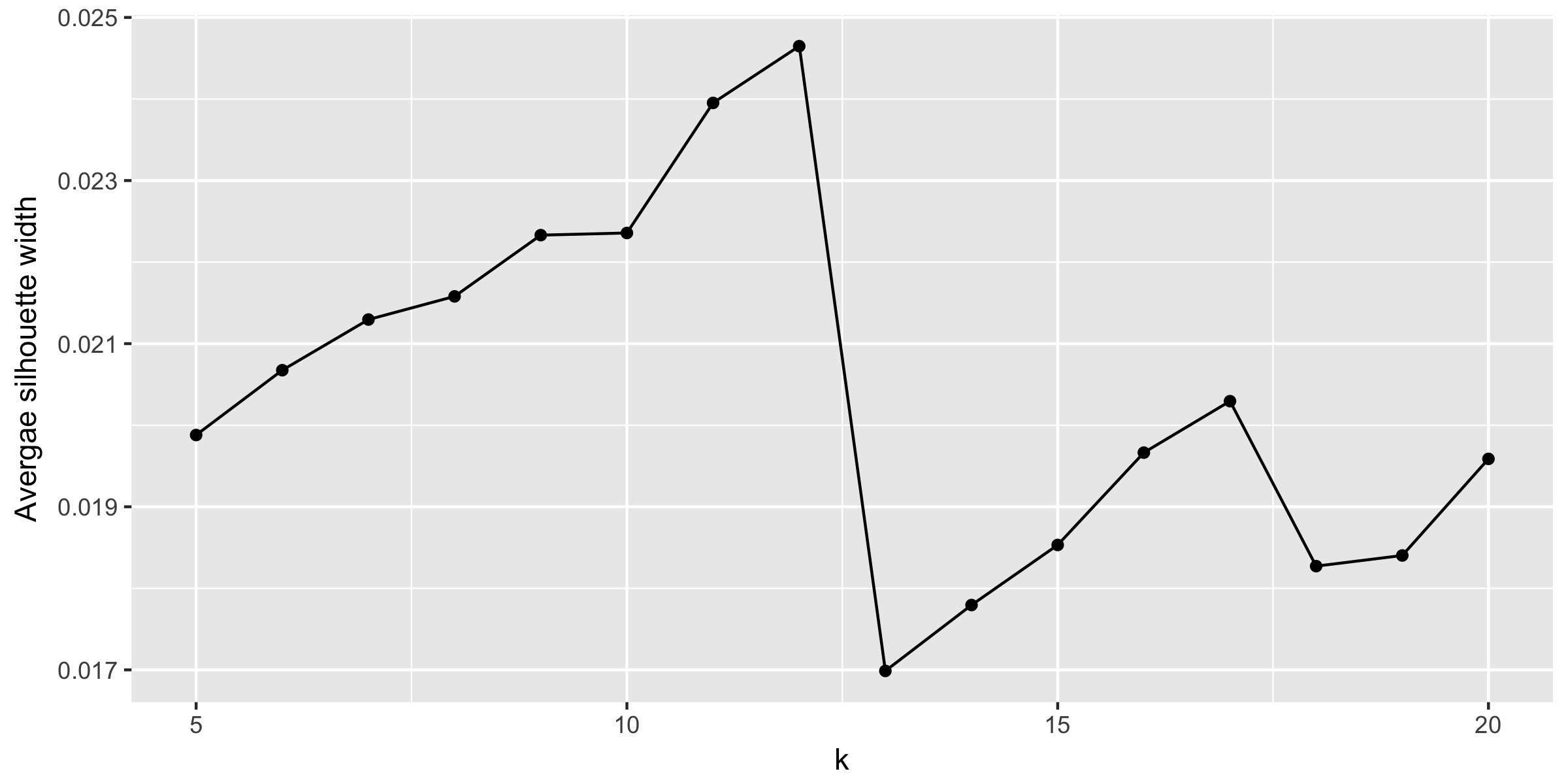}
\caption{Average Silhouette width based on 100 90\% subsamples of the set of word vectors.}
\label{fig:silhouette}
\end{figure}

\newpage
\section{APPENDIX: Cluster word clouds}
\label{sec:wordcloud}

\begin{figure}[H]
\centering
\includegraphics[scale = 0.42]{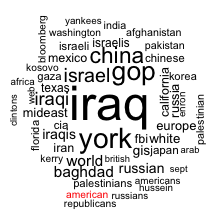}
\includegraphics[scale = 0.42]{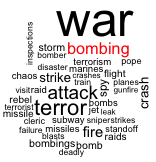}
\includegraphics[scale = 0.42]{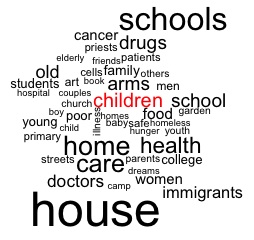}
\includegraphics[scale = 0.42]{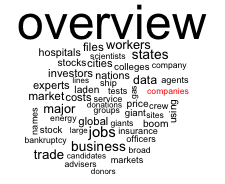}
\includegraphics[scale = 0.42]{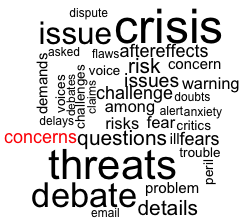}
\includegraphics[scale = 0.42]{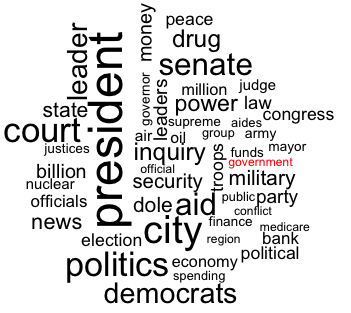}
\includegraphics[scale = 0.42]{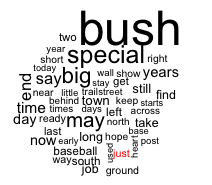}
\includegraphics[scale = 0.42]{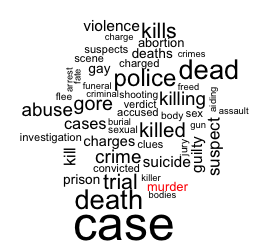}
\includegraphics[scale = 0.42]{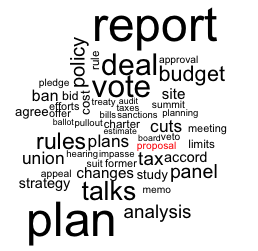}
\includegraphics[scale = 0.42]{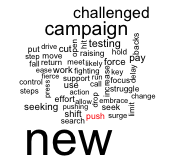}
\includegraphics[scale = 0.42]{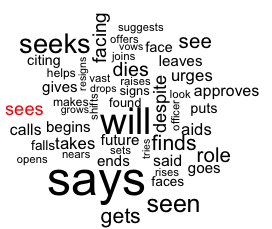}
\includegraphics[scale = 0.42]{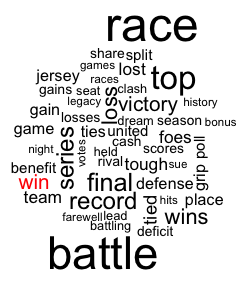}
\caption{Word clouds for the 12 word clusters. The word corresponding to the cluster center is highlighted in red. The size of each word corresponds to its frequency in the NY Times headlines corpus.}
\label{fig:wordclouds}
\end{figure}

\section{APPENDIX: Examples of images}
\label{sec:images}

\begin{figure}[H]
\centering
\includegraphics[scale = 0.45]{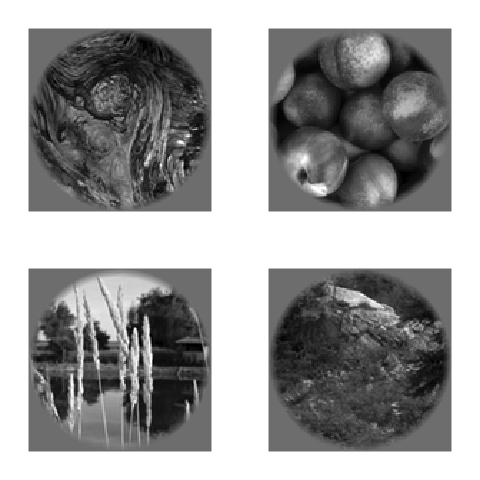}
\caption{Four randomly selected examples of validation images from the top cluster of images in Figure \ref{fig:all-voxel}.}
\label{fig:cl2}
\end{figure}

\begin{figure}[H]
\centering
\includegraphics[scale = 0.45]{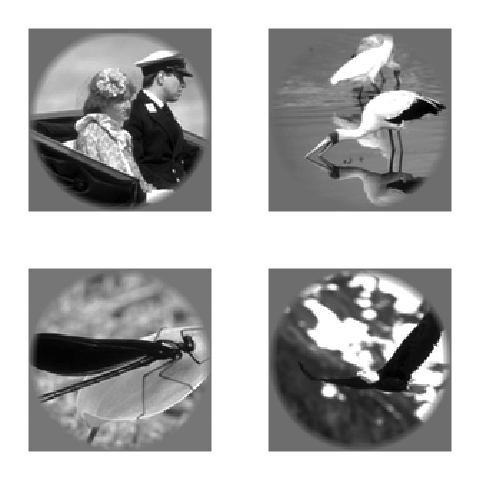}
\caption{Four randomly selected examples of validation images from the bottom cluster of images in Figure \ref{fig:all-voxel}.}
\label{fig:cl1}
\end{figure}

\newpage
\bigskip
\begin{center}
{\large\bf SUPPLEMENTARY MATERIAL}
\end{center}

\begin{description}

\item[R Markdown html documents:]  organ.html, word.html, fMRI.html: html documents generated by R Markdown that present the entire analysis pipeline for each of the three case studies.

\end{description}

\bigskip
\begin{center}
{\large\bf ACKNOWLEDGMENTS}
\end{center}

\noindent The authors would like to thank the Gallant Lab at UC Berkeley for providing the fMRI data. This research is partially supported by NSF grants DMS-1107000,  CDS\&E-MSS 1228246, DMS-1160319 (FRG), NHGRI grant 1U01HG007031-01 (ENCODE), AFOSR grant FA9550-14-1-0016, and the Center for Science of Information (CSoI), an USNSF Science and Technology Center, under grant agreement CCF-0939370.

\bibliography{Diagnostics}

\end{document}